\documentstyle[aps,preprint]{revtex}
\begin{document}
\draft
\title {COMPETITION BETWEEN QUASI-MOLECULAR RESONANCES AND FUSION-FISSION IN
LIGHT DINUCLEAR SYSTEMS} 

\author { C. Beck\thanks{ Presented at The International Conference ``Nuclear
Physics close to the Barrier", Warsaw, Poland, June 30 - July 4, 1998.
to be published in Acta Physica Polonica B, fall 1998}}

\address{\it Institut de Recherches Subatomiques, 
UMR7500, CNRS-IN2P3 et Universit\'e Louis Pasteur, 
23 rue du Loess - BP28,
F-67037 Strasbourg, France } 

\date{\today}
\maketitle

%\newpage

\begin{abstract}
{The results presented in this paper clearly suggest that a coherent framework
may exist which connects the topics of heavy-ion molecular resonances,
hyperdeformation effects, and fission shape isomerism. New data on
particle-particle-$\gamma$ triple coincidences of the $^{28}$Si+$^{28}$Si
reaction at a beam energy corresponding to the population of a conjectured
J$^{\pi}$ = 38$^{+}$ resonance in $^{56}$Ni are presented. The absence of
alignment of the spins of the outgoing fragments with respect to the orbital
angular momentum is found to be in contrast with the alignment as measured for
the $^{24}$Mg+$^{24}$Mg resonances. A molecular-model picture is presented to
suggest a ``butterfly" motion of two oblate $^{28}$Si nuclei interacting in a
equator-to-equator molecular configuration.} 
\end{abstract} 

\vskip 2.0cm
 
{ PACS numbers: 25.70.Jj, 25.70.Gh, 24.60.Dr }
% body of paper here

\newpage

The interplay between the reaction dynamics and the nuclear structure of the
interacting nuclei in light heavy-ion systems (A$_{cn}$ $\leq$ 60) has been the
focus of a number of experimental and theoretical studies over the last decade
[1-6]. It has been established that the fusion-fission (FF) process in such
light dinuclear systems has to be taken into account when exploring the
limitations of the complete fusion process at high excitation energies and
large angular momenta [2]. The detailed analysis of mass distributions and the
systematic investigation of different entrance channels populating a given
compound nucleus (CN) supports the FF picture. This is the case for the
$^{47}$V nucleus as formed in statistical way by three different reactions [3].
Another example is given in Fig.1 for the neighbouring $^{48}$Cr CN as
populated by two different entrance channels : $^{36}$Ar+$^{12}$C [4] and
$^{24}$Mg+$^{24}$Mg [5] the third reaction $^{32}$S+$^{24}$Mg [6] leads to the
$^{56}$Ni CN which will be further investigated in this paper. The data [4-6]
are plotted as solid points, whereas the histogramms are FF model predictions
as calculated by using the scission-point picture [7]. Transition-state model
(TSM) calculations using the saddle-point picture [6] give very similar
results. Typical excitation-energy spectra for the $^{32}$S+$^{24}$Mg reaction
are displayed for two bombarding energies in Fig.2 and well compared with TSM
calculations [6]. The FF mechanism is known to play a significant role at spins
above the grazing angular momentum so that the nuclear configuration leading to
the resonance behavior is only slightly more extended than that of the nuclear
scission (saddle) point. In this paper we will report on very recent results
showing for the first time that there exists a strong overlap between the FF
process and the quasi-molecular resonant behavior in the interaction between
two light nuclei.

Narrow-width resonances in excitation functions of $^{28}$Si~$+$~$^{28}$Si [8]
and $^{24}$Mg $+$ $^{24}$Mg [9] elastic and several inelastic scattering yields
were found to be correlated among these channels and they were believed to be
associated with quasi-stable configurations with extreme deformation. These
very striking quasi-molecular resonant structures are possibly connected to a
rather unusual subset of high-spin states stabilized against the mixing into
the more numerous CN states by some special symmetry. 
This interpretation is supported by theoretical investigations indicating that
shell-stabilized ``hyperdeformed" shapes may exist in the $^{56}$Ni and
$^{48}$Cr nuclei with large angular momenta [10]. Spin-alignment measurements
[11] for the resonant $\rm ^{24}Mg+{^{24}Mg}$ system [9] are available unable
us to suggest a deformed configuration (with an axis ratio of 3:0) for the
$^{48}$Cr compound system that corresponds to two prolate deformed $^{24}$Mg
nuclei in a pole-to-pole arrangement. Due to the complexity of the resonance
structure, where several narrow resonances are found to be described by the
same resonance spin, their decay properties are difficult to be analysed within
the Nilsson-Strutinsky approach [10] alone. This resonant behavior might be,
however, understood in the framework of a molecular picture developed by Uegaki
and Abe [12] using the strong-coupling limit. The motion of the two interacting
$^{24}$Mg nuclei in the $\rm ^{24}Mg+{^{24}Mg}$ collision is here described
with different vibrational excitations about the deformed equilibrium shape
leading to a fragmentation of the resonance strength for a given spin value.
Similar calculations have been done for the $\rm Si+Si$ system [13], however
with differences arising because of the oblate deformation of the $^{28}$Si
nucleus in its ground state. To explore these differences and to further
develop the properties of the $^{28}$Si+$^{28}$Si resonances, an experimental
study of the $^{28}$Si+$^{28}$Si collision at an energy corresponding to a
conjectured $J^{\pi}=38^{+}$ resonance in $^{56}$Ni [8] has been undertaken.
This study was motivated in part by the development of a new generation of
state-of-art 4$\pi$ $\gamma$-ray detector arrays that allows the high-precision
measurements needed to determine the resonance properties. 

Fragment-fragment-$\gamma$ triple coincidence events of the
$^{28}$Si~$+$~$^{28}$Si reaction have been measured [14] at the VIVITRON Tandem
facility at a resonance energy [8]. TSM model calculations [6] have been
performed in order to well separate the resonant yields from a strong FF
"background" as shown in Fig.3 for the excitation-energy spectrum of the
$^{28}$Si+$^{28}$Si exit-channel. The full spectroscopy [14] of the $^{28}$Si
exit-fragments shows that the resonant ``signal", as extracted from the Q-value
spectra, is related to a long-lived di-nuclear system stabilized by a favorable
oblate-oblate configuration before scission. Unexpected spin-alignment results
are shown in Fig.4 for the measured angular distributions (elastic, inelastic,
and mutual excitation channels) which are well described by a single Legendre
polynomial squared of order L = 38. 

Fragment-fragment-$\gamma$ coincidence data of the $^{28}$Si + $^{28}$Si
exit-channel have been measured for the low-lying excitation states (single
inelastic 2$^{+}_{1}$ and mutual inelastic (2$^{+}_{1}$, 2$^{+}_{1}$)
exit-channels) by using Eurogam Phase II. Three quantization axes have been
defined as follows : (a) corresponds to the beam axis, (b) axis normal to the
scattering plane, and (c) axis perpendicular to the (a) and (b) axes. The
fragment detectors are placed symmetrically with respect to the beam axis so
that the $^{28}$Si fragments are detected in the angular region 88$^{o}$ $\leq$
$\theta_{\rm c.m.}$ $\leq$ 92$^{o}$, then the c) axis corresponds
approximatively to the molecular axis parallel to the relative vector between
their two centers. The experimental results of the $\gamma$-ray angular
correlations for the mutual excitation exit-channel 2$^{+}$-2$^{+}$ are shown
in Fig.5 as solid points. The minima observed in a) and b) at 90$^{o}$ imply
that the intrinsic spin vectors of the 2$^{+}$ states lie in the reaction plane
and are perpendicular to the orbital angular momentum. The value of the angular
momentum remains close to L = 38 ${\rm \hbar}$ for the two exit channels, in
agreement with the fragment-fragment angular distributions of Fig.4. The
feeding of the $^{28}$Si states were also measured in this experiment [14]. The
study of the feeding of the bands of $^{28}$Si has revealed that the $^{28}$Si
is dominated by {\it oblate} deformation [14]. 

The prolate-prolate system $^{24}$Mg~$+$~$^{24}$Mg has been shown to be
characterised by spin alignment [11] in sharp contrast with the present results
of the oblate-oblate system $^{28}$Si~+~$^{28}$Si. Quasi-molecular resonances
in the $^{28}$Si~+~$^{28}$Si and $^{24}$Mg~$+$~$^{24}$Mg collisions are
examined in the framework of the molecular model [12,13]. The general ideas of
the molecular model are based on the following considerations : the occurence
of elongated but stable dinuclear configurations and their characteristic
normal-modes of motions at equilibrium may be linked to the
observed resonant structures. A stable configuration of the
$^{24}$Mg~$+$~$^{24}$Mg scattering is found to be a {\it pole-pole}
configuration, due to the prolate shape of the interacting nuclei $^{24}$Mg.
Around the equilibrium various dynamical modes which may appear could be
responsible for the observed narrow resonances. In an oblate-oblate system
$^{28}$Si~+~$^{28}$Si (the $^{28}$Si nucleus has an oblate shape with high
spins such as 38$\hbar$, an equator-equator touching configuration is favored.
In a Butterfly mode excitation, the interacting nuclei (fragments) move
coherently, therefore the intrinsic spins of nuclei are always anti-paralle l
to each other, i.e., the total intrinsic spin I = 0. On the other hand in an
anti-butterfly mode, the spins are parallel in order maximize the 
total spin. 

Figs.6 and 7 display molecular-model predictions on spin
distributions of the normal-mode motions for $^{28}$Si~+~$^{28}$Si and
$^{24}$Mg~$+$~$^{24}$Mg to the inelastic channel (2$_{1}^{+}$,2$_{1}^{+}$).
Fig.6 shows (see panel B) that there is a concentration of the probabilities
for channel spin of I = 0 for Butterfly motion. This is consistent with the
experimental angular distributions with L = \\
J in Fig.5. The curves in Fig.5 correspond to the $\gamma$-ray angular
distributions for the mutual inelastic channel of $^{28}$Si+$^{28}$Si obtained
by using the molecular-model calculations [15] including the K-mixing and the
``tilting" mode with J = L = 38 $\hbar$ (solid curve) and L = 34 $\hbar$
(dotted curve). The result of the 38 $\hbar$ curve exhibit strong concentration
in ``m = 0" states, which appears to be in fairly good qualitative agreement
with the experimental $\gamma$-ray distributions. The results of the
calculations of Fig.7 for $^{24}$Mg+$^{24}$Mg do not show a preference for the
Butterfly mode which is consistent with experimental observation of spin
alignment. 

Particle-particle-gamma coincident measurements in the decay of the
$^{28}$Si+$^{28}$Si 38$^{+}$ resonance [8] provide informations on the
orientations of the fragments spins in the resonance state being on the
reaction plane. The molecular model [13] which includes the K-mixing and to
give rise to a tilting mode assumes that the dinucleus configuration of
oblate-oblate systems is axially non-symmetric. This is different from
prolate-prolate cases such as $^{24}$Mg~$+$~$^{24}$Mg for which the pole-pole
configuration is axially symmetric. $K=0$ gives the isotropic distribution with
orientations of the fragments spins (pancakes) having the same distribution.
The angular distribution of gamma-ray emitted from the fragments is calculated
with this spin distribution and compared with the measurements in Fig.5. The
results exhibit strong concentration in "m = 0" states, which may be in good
agreement with the experimental $\gamma$-ray distribution. Work is in progress
[15] for a reasonable comparison between the data and the calculated results.
It seems that in the $^{28}$Si~+~$^{28}$Si system (oblate-oblate), the absence
of spin alignment could result from a Butterfly motion and tilting mode. 

The observation of spin alignment has been first made in $^{12}$C~$+$~$^{12}$C
quasi-molecular resonances [16]. In this case the correlation between the spin
orientations of the two $^{12}$C nuclei in the mutual inelastic scattering has
been deduced from the measured directional correlations of the
particle-coincident $\gamma$-ray. Resonances in the excitation functions were
found to be associated with the mutually aligned component. Two years after the
observation of the spin alignment in $^{12}$C~$+$~$^{12}$C molecular
resonances, A.~Wuosmaa and collaborators [11] have measured the single and
correlated magnetic-substate population parameters for 2$^{+}$ and
2$^{+}$-2$^{+}$ excitations in $^{24}$Mg~$+$~$^{24}$Mg (prolate-prolate system)
in the region of two strong resonances observed in inelastic scattering [7].
These data [11] have provided spectroscopic informations relatively
uncontaminated by nonresonant amplitudes, and allow spin assignment of
J$^{\pi}$ = 36$^{+}$ for two resonances E$_{\rm c.m.}$~=~45.70 and 46.65 MeV.
The angular correlation data for the mutual 2$^{+}$ inelastic scattering
channel suggest a dominant decay ${\rm \ell}$ value of ${\rm \ell}$ = 34 ${\rm
\hbar}$ for both resonances. Correlated spin alignment data for this channel
confirm the expectations for the relationship between angular momentum coupling
and spin alignment for these resonances. The relatively high spin values
suggest a resonance configuration in which the two $^{24}$Mg nuclei interact
pole-to-pole, allowing the system to sustain a large amount of angular
momentum. This results is well corroborated by the molecular model [12]. The
competition between quasi-molecular resonances and FF in the
$^{24}$Mg+$^{24}$Mg reaction was also investigated very carefully at Argonne
[4,5] and FF yields from the $^{28}$Si+$^{28}$Si reaction are presently being
measured at the VIVITRON facility [17]. 

The present fragment-fragment-$\gamma$ coincident data on the
$^{28}$Si~$+$~$^{28}$Si scattering as performed at the 38$^{+}$ resonance
energy shows for the first time the absence of spin alignment in a heavy-ion
collision. This has been primarly shown in the measured angular distributions
of the elastic, inelastic 2$^{+}$, and mutual excitation channels
2$^{+}$-2$^{+}$, which are dominated by a pure and unique partial wave with L =
38\ ${\rm \hbar}$, and has been confirmed by measuring their
particle-particle-$\gamma$ angular correlations. The non-resonant high-energy
structures in the experimental energy spectra are well described by a FF
picture as shown by Figs.2 and 3 for both the $^{32}$S+$^{24}$Mg and
$^{28}$Si+$^{28}$Si entrance channels which both populate the $^{56}$Ni CN.
Further experimental investigations (inclusive FF measurements as well as light
charged particle - fragments correlations measurements) of the fusion process
of the $^{28}$Si+$^{28}$Si reaction are underway at the VIVITRON facility using
the ICARE charged particle multi-detector array [17]. Within the molecular
model the lack of spin alignment is taken to suggest an excitation of the
Butterfly mode in the vibrational motion of the observed resonance. Therefore a
stable configuration is inferred to be an elongated one, namely an
equator-to-equator touching configuration. The comparison between the three
symmetric systems $^{12}$C~$+$~$^{12}$C, $^{24}$Mg~$+$~$^{24}$Mg and
$^{28}$Si~$+$~$^{28}$Si shows an interesting contrast in the spin orientation
at resonance energies. The $^{28}$Si~$+$~$^{28}$Si oblate-oblate system is
expected to be characterised with essentially no spin alignment in contrast to
the observed spin alignment for $^{12}$C~$+$~$^{12}$C oblate-oblate system
and $^{24}$Mg~$+$~$^{24}$Mg prolate-prolate system. The molecular-model
calculations explain the absence of alignment in the oblate-oblate
$^{28}$Si~$+$~$^{28}$Si system by a Butterfly motion. 

The question to understand why the orientations of spin in the two
oblate-oblate systems $^{28}$Si~$+$~$^{28}$Si and $^{12}$C~$+$~$^{12}$C are so
different and why the weak-coupling picture works well in the
$^{12}$C~$+$~$^{12}$C system is still open. On the other hand it is interesting
to observe how well the molecular model (strong-coupling picture) works for the
$^{28}$Si~$+$~$^{28}$Si dinuclear system. This is, of course, considered to be
due to a possible difference of interactions between the constituent nuclei.
Actually, the experiments show that in heavier systems, such as
$^{24}$Mg~$+$~$^{24}$Mg and $^{28}$Si~$+$~$^{28}$Si, the resonances are
observed to be correlated among many inelastic channels, suggesting that the
interaction is effectively strong and that the resonance states include many
components (partial widths) over reaction channels. The situation appears to be
different in the lighter dinuclear systems such as $^{12}$C~$+$~$^{12}$C. This
problem requires more systematic experimental and theoretical studies.
Experiments with the EUROBALL and GAMMASPHERE arrays are planned or already
accepted in order to measure precise excitation functions of the
$^{24}$Mg+$^{24}$Mg and $^{24}$Mg+$^{12}$C reactions which results will help in
developing the differences and possible relationships between the heavy-ion
resonance and compound-nucleus fission processes in light-mass dinuclear
systems. 

\vskip 1.0cm

\centerline{\bf Acknowledgements}

\bigskip

I am pleased to warmly acknowledge Profs. Y. Abe, T. Matsuse, S.J. Sanders, A.
Szanto de Toledo and E. Uegaki and my colleagues from IReS Strasbourg
especially R. Nouicer, F. Haas, V. Rauch, D. Mahboub, N. Aissaoui, C.
Bhattacharya and M. Rousseau for their contiuous help during the completion of
this study. Many thanks to the VIVITRON operators for providing us with well
focussed $^{28}$Si beams in both the EUROGAM and ICARE scattering chambers and
also to the EUROGAM Phase II staff of Strasbourg.

\begin{figure}
\caption{Experimental mass distributions (solid points) as measured for (a)
$^{36}$Ar+$^{12}$C at 188 MeV [4], (b) $^{24}$Mg+$^{24}$Mg at 88.8 MeV [5], (c)
$^{32}$S+$^{24}$S at 142 MeV [6] and compared to scission-point model
calculations (histogramms) using the Extended Hauser-Feshbach Method [7]. The
input parameters were taken from complete fusion data and from the FF
systematics [1,3,14].} 
\label{fig1}
\end{figure}

\begin{figure}
\caption{Excitation-energy spectra for the  $\rm ^{28}Si + {^{28}Si}$
exit-channel for the $^{32}$S+$^{24}$Mg reaction at (a) 119.1 MeV and (b) 129.0
MeV [1,6]. The curves correspond {\sc TSM} calculations [1] for fission decay
to particle-bound states (solid curve) and for all decays (dotted curve).} 
\label{fig2}
\end{figure} 

\begin{figure}
\caption{Excitation-energy spectrum for the $\rm ^{28}Si + {^{28}Si}$
exit-channel for the $^{28}$Si+$^{28}$Si reaction at 111.6 MeV [14]. Efficiency
corrected experimental data (solid points) are given as absolute cross
sections. The curve corresponds to {\sc TSM} calculations [1] for fission
decay.} 
\label{fig3}
\end{figure}

\begin{figure}
\caption{Experimental angular distributions of the elastic ($E_{\sc x} \approx 
0\rm\,MeV$), inelastic ($E_{\sc x} \approx 1.7\rm\,MeV$), mutual inelastic
($E_{\sc x} \approx 3.6\rm\,MeV$) and mutual excitations ($E_{\sc x} \approx 
5.7$ and  $8.7\rm\,MeV$) for the $\rm ^{28}Si + {^{28}Si}$ symmetric
exit-channel [14]. The solid curves represent squared Legendre polynomial of
order 38. The dotted curve corresponds to a 1/$\sin \theta_{\rm c.m.}$
behavior.} 
\label{fig4}
\end{figure}

\begin{figure}
\caption{Experimental $\gamma$-ray angular correlations of the mutual inelastic
states ($2^{+}_{1}$,~$2^{+}_{1}$) in the angular region $88^{\circ} \le
\theta_{\rm c.m.}^{F}({\rm ^{28}Si}) \le 92^{\circ}$ of the $\rm ^{28}Si +
{^{28}Si}$ exit-channel [14] for three quantization axes as defined in the
text. The curves are molecular-model predictions discussed in the text.} 
\label{fig5}
\end{figure}

\begin{figure}
\caption{Molecular-model predictions [15] for the probability distributions of 
$\rm ^{28}Si + {^{28}Si}$ ([ 2$^{+}_{1} \otimes $ 2$^{+}_{1}$] with $L = J -
I$) system versus channel spin $I$.} 
\label{fig6}
\end{figure}

\begin{figure}
\caption{Molecular-model predictions [15] for the probability distributions of 
$\rm ^{24}Mg + {^{24}Mg}$ ([ 2$^{+}_{1} \otimes $ 2$^{+}_{1}$] with $L = J -
I$) system versus channel spin $I$.} 
\label{fig7}
\end{figure}

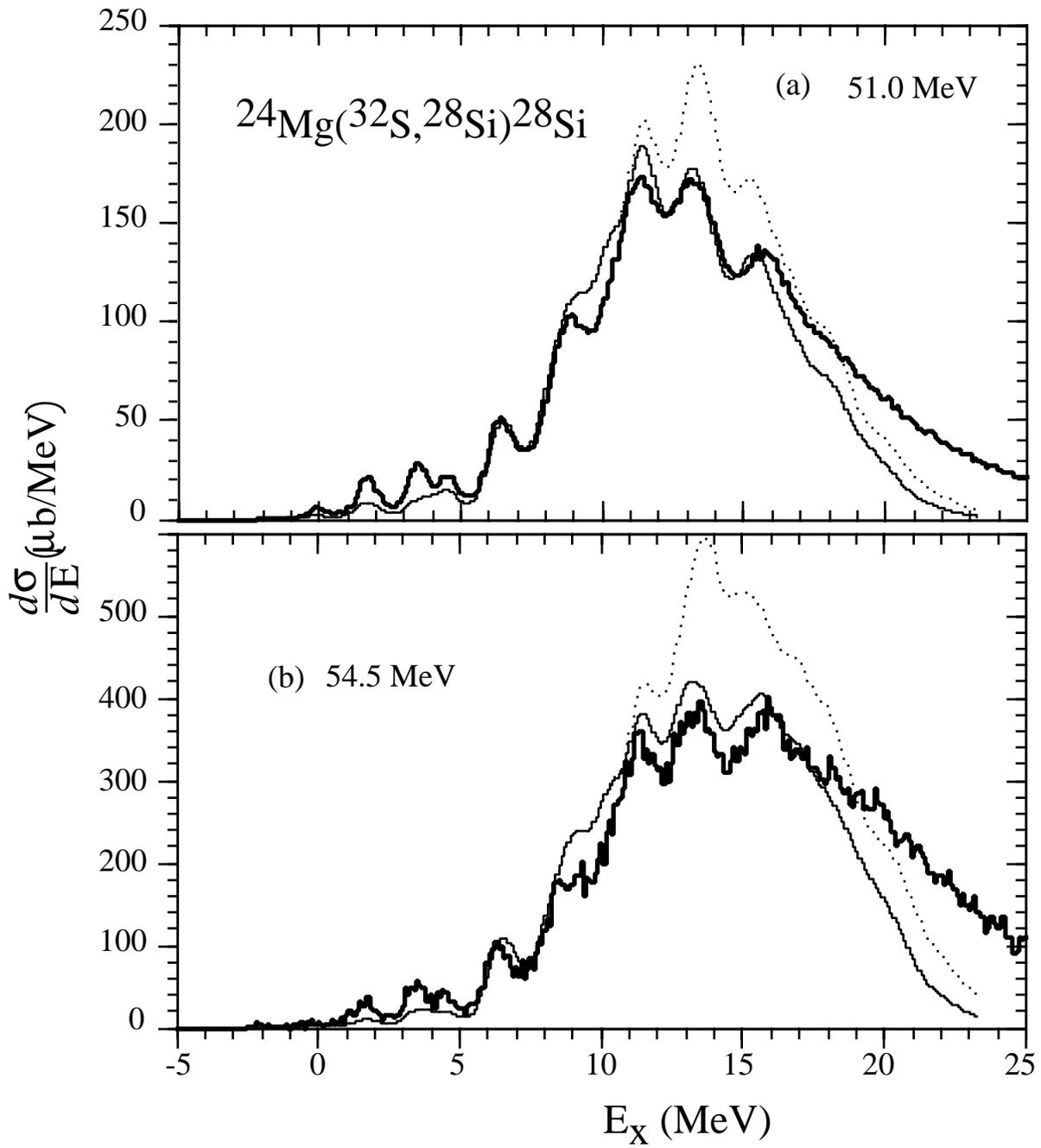
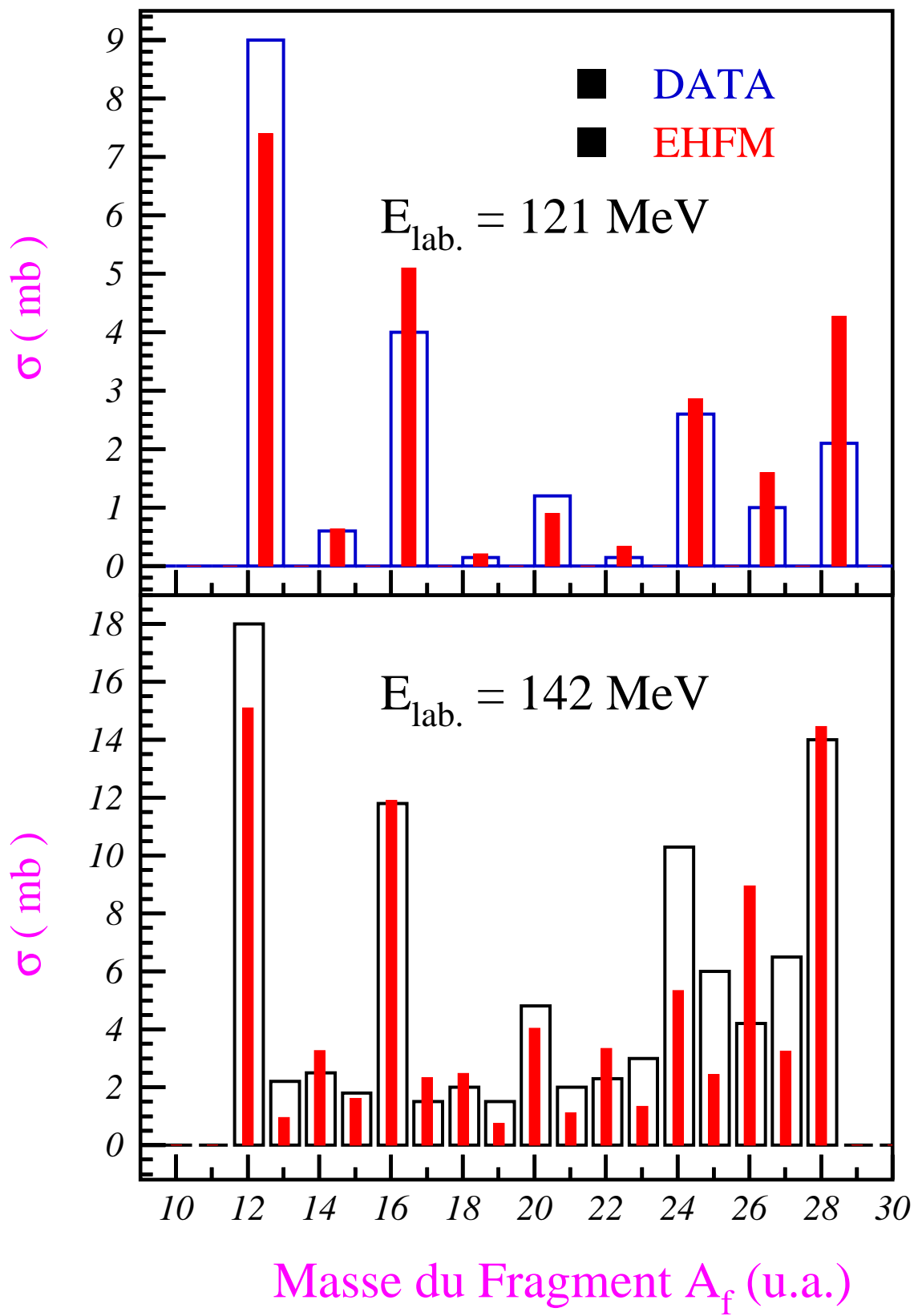
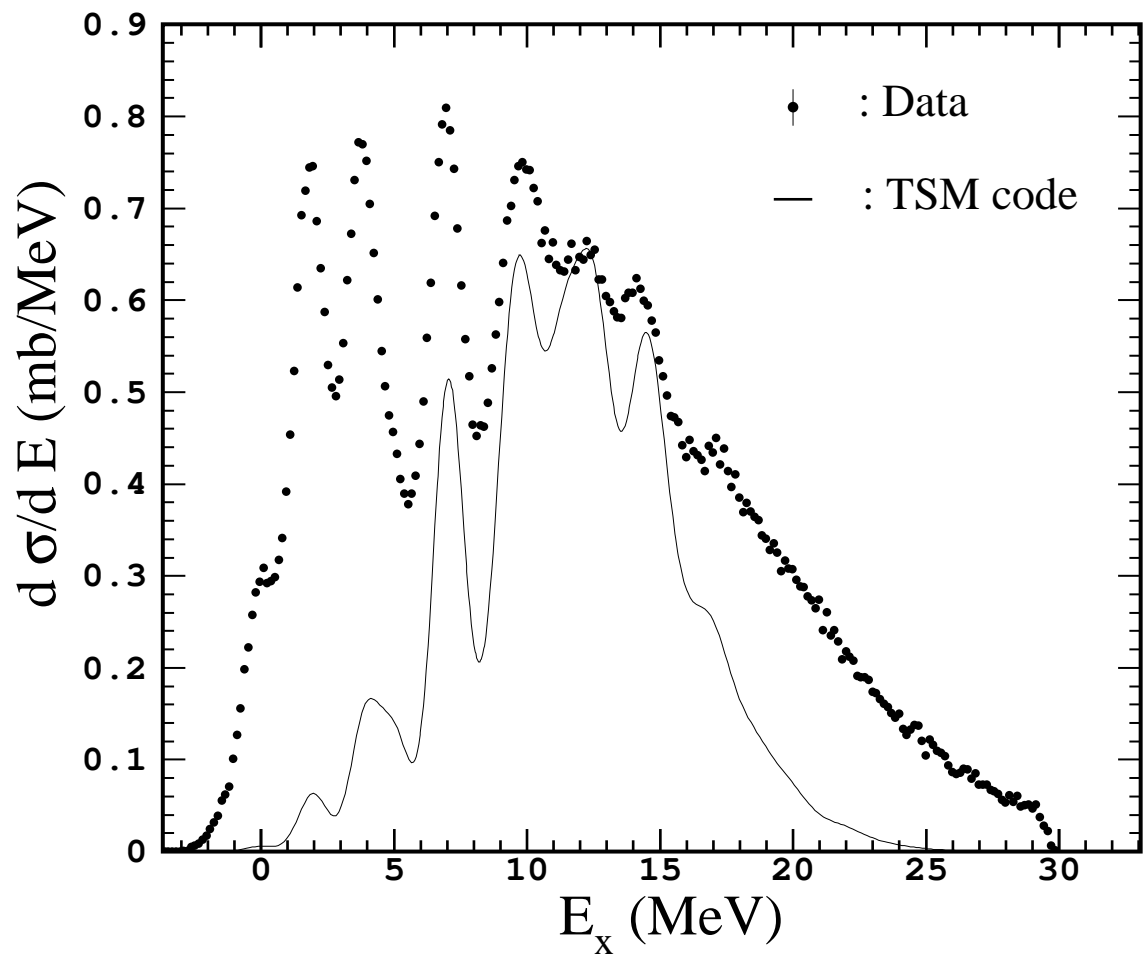
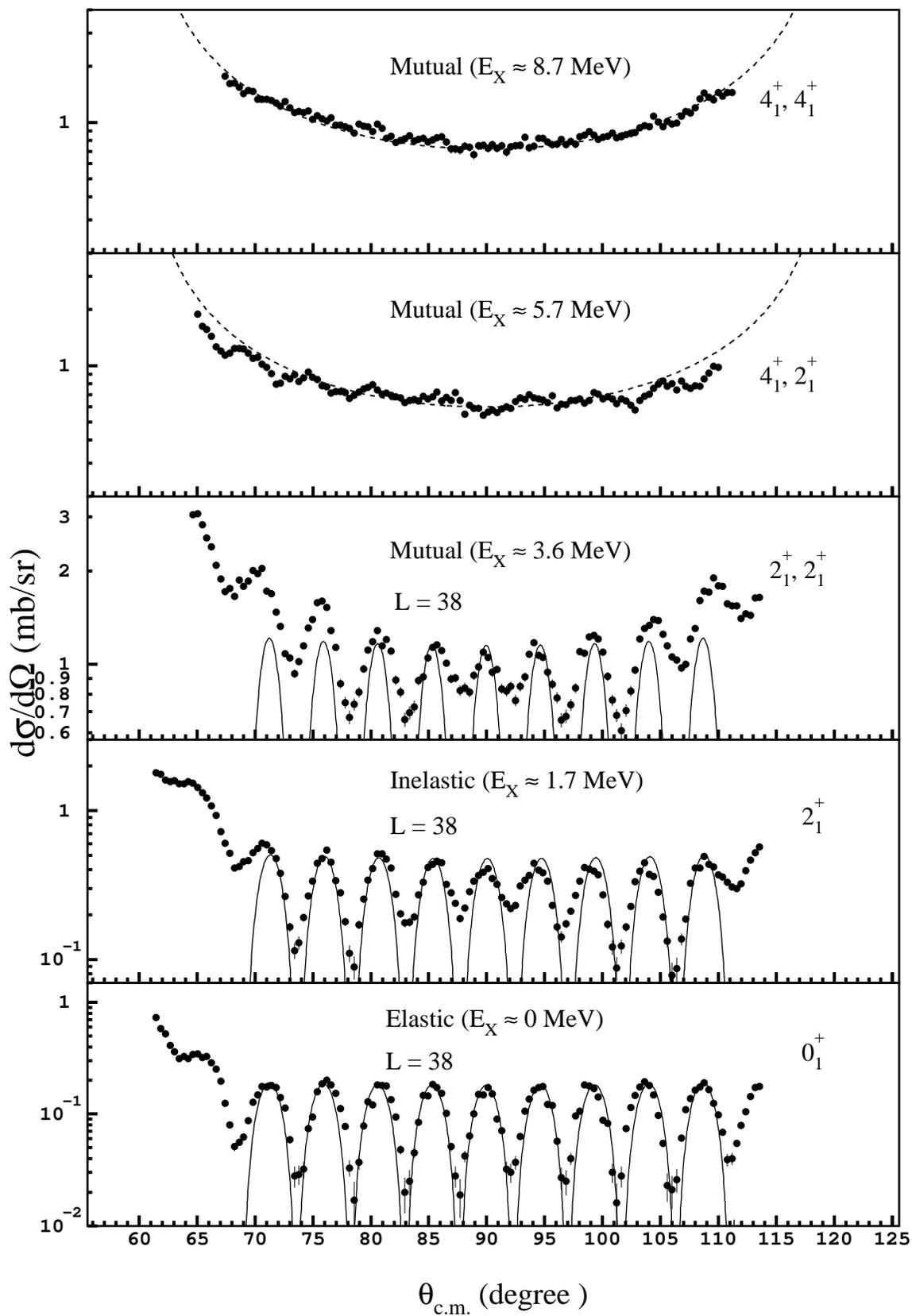
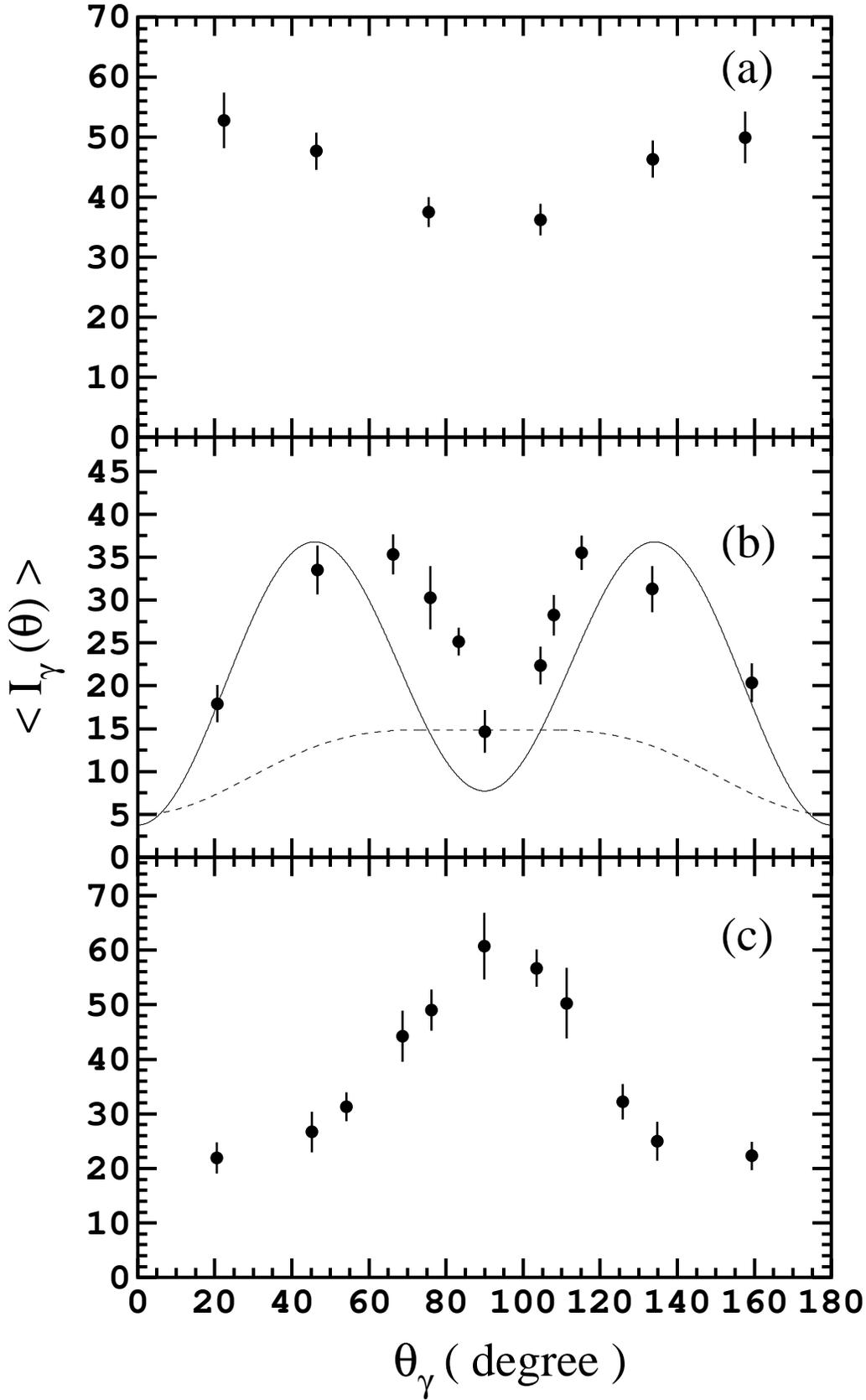
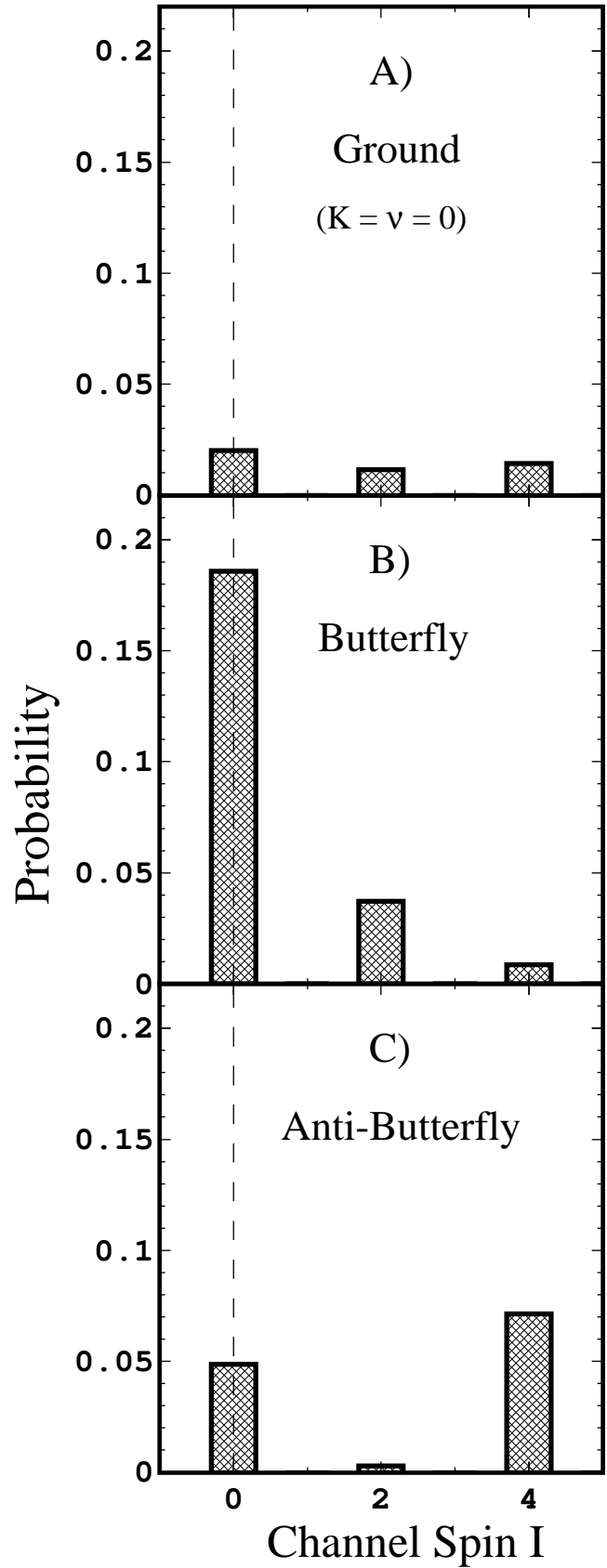
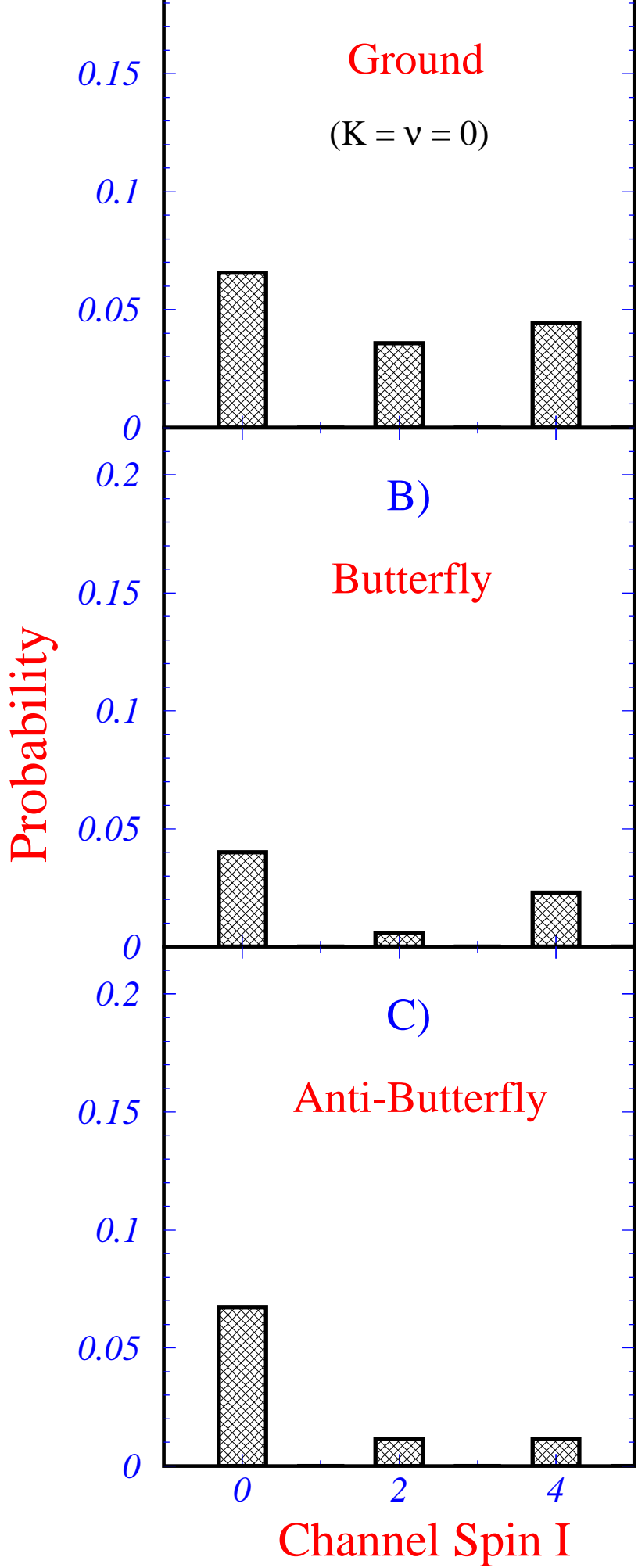
\end{document}